\documentclass[showpacs,prl,preprintnumbers,amsmath,amssymb]{revtex4}

\usepackage{graphicx}
\usepackage{dcolumn}
\usepackage{bm}

\begin{document}

\title{Thermodynamics induced geometry of self - gravitating systems}

\author{B. I. Lev }

\affiliation{Bogolyubov Institute for Theoretical Physics, NAS of
Ukraine, Metrolohichna 14-b, Kyiv 03680, Ukraine, e-mail bohdan.lev@gmail.com }

\date{\today}

\begin{abstract} 
\textbf{Abstract} A new approach based on a statistical operator is presented, which allows to take into account the inhomogeneous particle distribution induced by gravitational interaction.
This method uses the saddle point procedure to find the dominant contribution to the statistical sum and allows to obtain all thermodynamic relations of self-gravitating systems. Based on thermodynamic relations, a description of thermodynamically induced geometry of matter distribution was proposed. Equations corresponding to the extremum of the statistical sum completely reproduce the well-known equations of the general theory of relativity.

\end{abstract}

\maketitle

\textbf{Introduction}

 The study of interacting particles systems provides, along with other advantages, the development and testing of fundamental ideas of statistical mechanics and thermodynamics. A few model systems of interacting particles are known, as far as the partition function can be exactly evaluated, at least, in the thermodynamic limit. Particle systems with long range interaction sometimes cannot be described in terms of a usual thermodynamic ensemble \cite{Tir}, \cite{Cav} and hence cannot be treated by standard methods of equilibrium statistical mechanics. In particular, inasmuch as energy is non-additive, the canonical ensemble is inapplicable to the study of systems with long-range interactions. Since equilibrium states are only local entropy maximum \cite{Zub}.The system is stable but thermodynamic limit does not exist \cite{Lal}. It is generally believed that for such systems the mean-field theory is exact. In this approach any thermodynamic function depends on the parameters only in terms of dimensionless combinations. 

Systems with long-range interactions, e.g., self-gravitating systems do not relax towards usual Boltzmann-Gibbs thermodynamic equilibrium. They get trapped in quasi-stationary states whose lifetimes diverge as the number of particles increases. A quantitative description of the instability threshold of spontaneous symmetry breaking for d-dimensional systems is given in Ref. \cite{Pak}. The homogeneous particle distribution in interacting systems is unstable and spatially inhomogeneous appear from the beginning. The behavior of interacting systems with various equilibrium ensembles should be described in different ways, mainly because any state of an interacting systems is far from equilibrium and the time of relaxation towards equilibrium state is very long. Nonequilibrium stationary states was described in Ref. \cite{Ben} and was shown, that three-dimensional systems to be trapped in  quasi- stationary states rather than evolve towards thermodynamic equilibrium. 

In this article we develop a new approach based on the non-equilibrium statistical operator \cite{Zub} that is more suitable for the treatment of systems of interacting particles. The equation of state and all thermodynamic characteristics needed are governed by the equations which contribute mainly to the partition function. The behavior of the systems is due to the appropriate thermodynamic relations. The main idea of article is to provide a detailed description of self -gravitating systems based on the principles of non -equilibrium statistical operator \cite{Zub}, and obtain a statistical justification by the gravitational equation general relativity. An important result of this approach is the determination of all the necessary thermodynamic relations for particle systems and the statistical induced dynamics of systems. As a result, we can present a proof of the general theory of relativity from the thermodynamic principle, which is realized by this statistical description. The purpose of this paper is to find the effective space of a thermodynamically stable distribution of self - gravitating systems and to give a geometrical description  induced by the character and intensity of the interaction.

\textbf{Statistical description self-gravitating systems}

Statistical mechanics based on conservation laws of the averages value dynamic variables. For the determination thermodynamic functions of systems are need use the presentation of corresponding statistical ensembles which take into account the  all states of this systems. In this case can determine ensemble as totality of system which be contained in same stationary external action. This systems have same character of contact with thermostat and possess all possible value macroscopic parameters which compatibility the necessary conditions. In systems, which are in same stationary external condition will are formed local equilibrium stationary distribution. If external condition will be depend from time that local equilibrium distribution are not stationary. For exactly determination local equilibrium ensemble must accordingly determine the distribution function or statistical operator of the system \cite{Zub}. Finally, can recall that the stable states of equilibrium of classical particles are only metastable because they correspond to local maximal of entropy from which can determine behavior of the systems. For now, let us briefly recall the approach \cite{Zub} that was developed in the articles \cite{Lev1}-\cite{Lev3} to describe the spatially inhomogeneous distribution of interacting particles.

If assume that non-equilibrium states of systems can determine through distribution energy $H(\mathbf{r})$ and number of particles (density) 
$n\left(\mathbf{r}\right)$ the  statistical operator local equilibrium distribution can write in the form \cite{Zub}:
\begin{equation}
Q_{l}=\int D\Gamma \exp\left\{-\int
\left(\beta(\mathbf{r})H(\mathbf{r})-\eta(\mathbf{r})n(\mathbf{r})\right)d\mathbf{r}\right\}
\end{equation} 
The integration in present formula must take over all phase space of system.
Must note, that in the case local equilibrium distribution Lagrange multipliers 
$\beta(\mathbf{r})$ and $\eta(\mathbf{r})$ are function of spatial point. The 
microscopic density of particles can present in standard form:
\begin{equation}
n(\mathbf{r})=\sum_{i}\delta(\mathbf{r}-\mathbf{r_{i}})
\end{equation} 
The introduction local equilibrium distribution are possible if relaxation time 
in all systems is more as relaxation time on local macroscopically area as part of 
this systems. The conservation number of particles and energy in systems can present in form natural relations $\int n(\mathbf{r})d\mathbf{r}=N$ and $ \int H(\mathbf{r})d \mathbf{r}=E$. After determination of non-equilibrium statistical operator can obtain all thermodynamic parameter nonequilibrium systems. Phenomenological thermodynamic based on the conservation lows for average value of 
physical parameter as number of particles, energy and momentum . For this can determine thermodynamic relation for inhomogeneous systems. The variation of statistical operator by Lagrange multipliers can write necessary thermodynamic relation in the form \cite{Zub}:
\begin{equation}
-\frac{\delta \ln Q_{l}}{\delta \beta(\mathbf{r})}=\langle
H(\mathbf{r})\rangle_{l}=E;    \frac{\delta \ln Q_{l}}{\delta \eta (\mathbf{r})}=\left\langle
n(\mathbf{r})\right\rangle_{l}=N
\end{equation} 
This relation is natural general prolongation well-known relation which take place in the case equilibrium systems, on the case inhomogeneous system for fixed energy $E$ and number of particles $N$.

For further statistical description of states is necessary determine Hamiltonian of system. In general relativistic case Hamiltonian of systems of interacting particle can present as:
\begin{equation}
H=\sum_{i}\frac{m_{i} c^{2}}{\sqrt{1-\frac{v^{2}_{i}}{c^{2}}}}-\frac{1}{2}\sum_{i,j} W(\mathbf{r_{i}}\mathbf{r_{j}})
\end{equation}
where $W(\mathbf{r_{i}}\mathbf{r_{j}})$ determine the attractive gravitational interaction and $v_{i}$ velocity of particles. Energy density of systems can  write in the following form:
\begin{equation}
H(\mathbf{r})= \Big(m(\mathbf{r}) c^{2} +\frac{p^{2}(\mathbf{r})}{2m(\mathbf{r})}\Big) n(\mathbf{r})-\frac{1}{2}\int
W(\mathbf{r},\mathbf{r'})n(\mathbf{r})n(\mathbf{r'})d \mathbf{r'}
\end{equation}
for small velocity of particles. This density energy of a systems is possible to use if we smash to equal bits all space with equal mass and consider moving in phase space not compressed fluid. In our case of system with gravitational character of interaction can write nonequilibrium statistical operator in the form:
\begin{equation}
Q_{l}=\int D\Gamma \exp\left\{-\int
\left(\beta(\mathbf{r})\frac{p^{2}(\mathbf{r})}{2m(\mathbf{r})}+\beta m(\mathbf{r}) c^{2} -\eta(\mathbf{r})\right)n(\mathbf{r})d\mathbf{r}+\frac{1}{2}\int
\beta(\mathbf{r})W(\mathbf{r},\mathbf{r'})n(\mathbf{r})n(\mathbf{r'}) d \mathbf{r} d \mathbf{r'} \right\}
\end{equation}
The integration over phase space can present as $D\Gamma=\frac{1}{\left( 2\pi \hbar\right)^{3}}\prod\limits_{i}dr_{i} dp_{i}$. 

In order to perform a formal integration in second part of this presentation, 
additional field variables can be introduced making use of the theory of Gaussian 
integrals \cite{Kle} - \cite{Str} :
\begin{equation}
\exp \left\{ -\frac{\nu ^2}{2}\int
\beta(\mathbf{r})\omega(\mathbf{r},\mathbf{r'})n(\mathbf{r})n(\mathbf{r'})d
\mathbf{r} d \mathbf{r'}\right\} = \int D \sigma \exp \left\{
-\frac{\nu ^2}{2}\int
\omega(\mathbf{r})^{-1}\sigma(\mathbf{r})\sigma(\mathbf{r'})d
\mathbf{r}d \mathbf{r'}-\nu \int\sqrt{\beta(\mathbf{r})}\sigma(\mathbf{r})n(\mathbf{r})d
\mathbf{r} \right\}
\end{equation}
where $D\sigma=\frac{\prod\limits_{s}d\sigma _{s} }{\sqrt{\det 2\pi \beta \omega(\mathbf{r},\mathbf{r'})}}$ and $\omega^{-1}(\mathbf{r},\mathbf{r'})$ is the inverse operator which satisfies the condition
$\omega^{-1}(\mathbf{r},\mathbf{r'})\omega(\mathbf{r'},\mathbf{r''})=\delta
(\mathbf{r}-\mathbf{r''})$, that means: the interaction energy presented the Green 
function for this operator and $\nu ^2=\pm 1$ depending on the sign of the interaction 
or the potential energy. After present manipulation the field variable 
$\sigma(\mathbf{r}) $ contains the same information as original distribution 
function, i.e. all information about possible spatial states of the systems. 

After this manipulation the statistical operator can rewrite in the form:
\begin{equation}
Q_{l}=\int D\Gamma \int D\varphi \exp\left\{-\int
\left(\beta(\mathbf{r})\frac{p^{2}(\mathbf{r})}{2m(\mathbf{r})}+\beta(\mathbf{r}) m(\mathbf{r}) c^{2} -\eta(\mathbf{r})-\sqrt{\beta(\mathbf{r}))}\varphi(\mathbf{r})\right)n(\mathbf{r})d\mathbf{r}-
\right\}Q_{int}
\end{equation}
where part which come from interaction
\begin{equation}
Q_{int}=\exp\left\{\frac{1}{2}\int 
W(\mathbf{r},\mathbf{r'})^{-1}\varphi(\mathbf{r})\varphi(\mathbf{r'})d\mathbf{r}d \mathbf{r'}\right\}
\end{equation}
In this general functional integral can be provide the integration on phase space. 
If use the definition of density we can rewrite the non-equilibrium statistical 
operator as:
\begin{equation}
Q_{l}=\int D\varphi \int \frac{1}{\left( 2\pi \hbar\right)
	^{3}N!}\prod\limits_{i}dr_{i} dp_{i}\xi(\mathbf{r_{i}})\exp\left\{-
\left(\beta(\mathbf{r_{i}})\frac{p_{i}^{2}}{2m_{i}}+\beta(\mathbf{r_{i}}) m(\mathbf{r_{i}}) c^{2}-\sqrt{\beta(\mathbf{r_{i}}))}\varphi(\mathbf{r_{i}}\right)\right\}Q_{int}
\end{equation}
where introduce the new variable $\xi(\mathbf{r})\equiv \exp \eta(\mathbf{r})$ 
which can interpreter as chemical activity. Now can make integration over impulse. 
The non-equilibrium statistical operator take the form \cite{Lev,Lev1}:
\begin{equation}
Q_{l}=\int D\varphi  Q_{int} \sum_{N}\frac{1}{N!}\left({\int d\mathbf{r} \xi(\mathbf{r})\left(\frac{2\pi m_{i}}{\hbar^{3}\beta(\mathbf{r})}
	\right)^{\frac{3}{2}}\exp \left(\beta(\mathbf{r_{i}}) m(\mathbf{r_{i}}) c^{2}+\sqrt{\beta(\mathbf{r})}\varphi(\mathbf{r})\right)}\right)^{N}
\end{equation}
After that the non-equilibrium statistical operator can be present as:
\begin{equation}
Q_{l}=\int D\varphi  Q_{int} \exp\left\{\int \left[
\xi(\mathbf{r})\left(\frac{2\pi m(\mathbf{r})}{\hbar^{3}\beta(\mathbf{r})}
\right)^{\frac{3}{2}}\exp\left(\beta(\mathbf{r}) m(\mathbf{r}) c^{2}+\sqrt{\beta(\mathbf{r})}\varphi(\mathbf{r})\right)\right]d\mathbf{r}\right\}
\end{equation}
We can present the nonequilibrium statistical operator in the term additional fields as:
\begin{equation}
Q_{l}=\int D\varphi  d \xi
\exp -\left\{S(\varphi(\mathbf{r}),\xi(\mathbf{r}),\beta(\mathbf{r}))\right\}
\end{equation}
where effective nonequilibrium "local thermodynamic potential" take the form:
\begin{equation}
\begin{tabular}{l}
$S=\frac{1}{2}\int
U(\mathbf{r},\mathbf{r'})^{-1}\varphi(\mathbf{r})\varphi(\mathbf{r'})d\mathbf{r}d \mathbf{r'}$ \\ 
$-\int \left[
\xi(\mathbf{r})\left(\frac{2\pi m(\mathbf{r})}{\hbar^{3}\beta(\mathbf{r})}
\right)^{\frac{3}{2}}\exp\left(\beta(\mathbf{r}) m(\mathbf{r}) c^{2}+\sqrt{\beta(\mathbf{r})}\varphi(\mathbf{r})\right)\right]d\mathbf{r}$ \\ 
\end{tabular} \label{Eq}
\end{equation}
The functional $S(\varphi(\mathbf{r}), \xi(\mathbf{r}),\beta(r))$ depends on the distribution of the field variables $\varphi(\mathbf{r}) $ and $\psi(\mathbf{r})$ which describe repulsive and attractive interaction, the chemical activity $\xi(\mathbf{r})$ and inverse temperature $\beta(\mathbf{r})$. The statistical operator allows obtain use the of efficient methods developed in the quantum field theory without imposing additional restrictions of integration over field variables or the perturbation theory. The saddle point method can now be further employed to find the asymptotic value of the statistical operator $Q_{l}$ for increasing number of particles $N$ to $\infty $. The dominant contribution is given by the states which satisfy the extreme condition for the functional. It's easy to see that saddle point equation present thermodynamic relation and it can write in the other form as: equation for field variables: $\frac{\delta S}{\delta \varphi(\mathbf{r})}=0$ for the normalization condition $\frac{\delta S}{\delta (\eta(\mathbf{r}))}=-\int \frac{\delta S}{\delta \xi(\mathbf{r})}\xi(\mathbf{r})d\mathbf{r}=N $ and for the conservation the energy of the systems $-\int \frac{\delta S}{\delta \beta (\mathbf{r})}d\mathbf{r}=E$. Solution of this equation fully determine all macroscopic thermodynamic parameters and describe general behavior of interacting system, whether this distribution of particles is spatially inhomogeneous or not. The above set of equations in principle solves the many-particle problem in thermodynamic limit. The spatially inhomogeneous solution of this equations correspondent the distribution of the interacting particles. Such inhomogeneous behavior is associated with the nature and intensity of the interaction. In other words, accumulation of particles in a finite spatial region (formation of a cluster) reflects the spatial distribution of the fields, the activity and temperature. Very important note, that only in this approach can take into account the inhomogeneous distribution temperature and chemical potential, which can depend from spatial distribution of particles in system.

\textbf{Thermodynamic relation}

For farther consideration we will be analyze the general presentation of "local thermodynamic potential" \eqref{Eq}.  If introduce the new field variables as $\sqrt{\beta(\mathbf{r}))}\varphi(\mathbf{r}=\Phi$  we can re-wright  \eqref{Eq} in the the more simple form:  
\begin{equation}
S=\frac{1}{2}\int
(\beta(\mathbf{r})W(\mathbf{r},\mathbf{r'}))^{-1}\Phi(\mathbf{r})\Phi(\mathbf{r'})d\mathbf{r}d \mathbf{r'}-\int \xi(\mathbf{r})\Lambda^{-3} \exp (\beta(\mathbf{r}) m(\mathbf{r}) c^{2}+\Phi(\mathbf{r}))d\mathbf{r}
\end{equation}
where introduce the definition local thermal de-Broglie wavelength $\Lambda(\mathbf{r})=\left(\frac{\hbar^{2}\beta(\mathbf{r})}{2 m(\mathbf{r})}
\right)^{\frac {1}{2}} $. To draw more information about the behavior of interacting systems, we also introduce some new variable. From normalization condition \eqref{NC} we can obtain 
\begin{equation}
\int \xi(\mathbf{r})\Lambda^{-3} \exp (\beta(\mathbf{r}) m(\mathbf{r}) c^{2}+\Phi(\mathbf{r}))d\mathbf{r}=N \label{NC}
\end{equation}
provides the macroscopic density  whose definition is given by
\begin{equation}
\rho(\mathbf{r})\equiv \xi(\mathbf{r})\Lambda^{-3} \exp (\beta(\mathbf{r}) m(\mathbf{r}) c^{2}+\Phi(\mathbf{r}))d\mathbf{r} \label{Den}
\end{equation}
In the case without interaction (for free particles $ \Phi(\mathbf{r})=0 $, if write the chemical activity in terms of the chemical potential $\xi(\mathbf{r})=exp (\mu(\mathbf{r})\beta(\mathbf{r}))$, 
can obtain the relation $\beta(\mathbf{r})\mu(\mathbf{r})=\beta(\mathbf{r}) m(\mathbf{r}) c^{2}+\ln \rho(\mathbf{r})\Lambda^{3}(\mathbf{r})$ that generalizes the relation in the equilibrium statistical mechanics for relativistic systems.

From minimum "local thermodynamic potential in the term new variables we obtain the next equation:
\begin{equation}
\int(\beta(\mathbf{r}) W(\mathbf{r},\mathbf{r'}))^{-1}\Phi(\mathbf{r'})d \mathbf{r'}+\rho(\mathbf{r})=0 
\end{equation}
If multiple such equation on $\int W(\mathbf{r},\mathbf{r'})d \mathbf{r'}$ we can obtain the equation from which can definition of the field variables. 
\begin{equation}
\Phi(\mathbf{r})+\int \beta(\mathbf{r}) W(\mathbf{r},\mathbf{r'})\rho(\mathbf{r'}) d \mathbf{r'}=0 \label{f}
\end{equation}
After that, in the case of relativistic particles with gravitational interaction from presentation \ref{Den} we can wright the general determination of chemical potential:
\begin{equation}
\beta(\mathbf{r})\mu(\mathbf{r})=\beta(\mathbf{r}) m(\mathbf{r}) c^{2}+\Phi+ \ln \rho(\mathbf{r})\Lambda^{3}(\mathbf{r})=\beta(\mathbf{r}) m(\mathbf{r}) c^{2}+\int \beta(\mathbf{r}) W(\mathbf{r},\mathbf{r'})\rho(\mathbf{r'}) d \mathbf{r'}+\ln \rho(\mathbf{r})\Lambda^{3}(\mathbf{r})
\end{equation}
Now we can determine the average value of the energy of systems using thermodynamic relation:
\begin{equation}
\langle H \rangle =-\int \frac{\delta \ln Q_{l}}{\delta \xi(\mathbf{r})}\frac{\delta\xi}{\delta \beta(\mathbf{r})}\rho(\mathbf{r})d \mathbf{r}=\int \mu(\mathbf{r})\rho(\mathbf{r})d\mathbf{r}
\end{equation}
that transform to general definition:
\begin{equation}
E=\int \rho(\mathbf{r})\langle H \rangle d\mathbf{r} =\int m(\mathbf{r}) c^{2}\rho(\mathbf{r})d\mathbf{r} +\int W(\mathbf{r},\mathbf{r'})\rho(\mathbf{r})\rho(\mathbf{r'}) d \mathbf{r'} +\int kT\rho(\mathbf{r})\ln \rho(\mathbf{r})\Lambda^{3}(\mathbf{r})d\mathbf{r}
\end{equation}
The last part of this relation exact equal to entropy of systems that lead to determination free energy of systems \cite{Fey}
\begin{equation}
F=\int m(\mathbf{r}) c^{2}\rho(\mathbf{r})d\mathbf{r} +\int W(\mathbf{r},\mathbf{r'})\rho(\mathbf{r})\rho(\mathbf{r'}) d \mathbf{r'}\label{FERT}
\end{equation}
After that we have all necessary thermodynamic relation for determination dynamic  behavior of non-equilibrium system \cite{Lif}, \cite{Gor}.  Such presentation lead determine evolution of non-equilibrium system in general case, but that is possible describe only in spacial case \cite{Lev,Lev1,Lev3,Lev4}. 

\textbf{Statistical induced the geometry of self-gravitating systems}

Any physical theory is based on the postulated geometric properties of the space. The problem of the geometry as a whole is equivalent to the problem of the behavior of the fields which form the space \cite{Hilb}- \cite{Wu}. Next we propose a geometric description of a thermodynamically stable distribution of self- interacting systems. The character and intensity of interaction in systems determine the effective geometry of the medium which is provided by the minimum of the total free energy \cite{Rebesh}.
Firs from all can remember that after Wick transformation integration of free energy by time reduced to action in Minkowski space
\begin{equation}
S=\frac{1}{c}\int m(\mathbf{r}) c^{2}\rho(\mathbf{r})d\mathbf{r} d ct +\frac{1}{c} \int W(\mathbf{r},\mathbf{r'})\rho(\mathbf{r})\rho(\mathbf{r'}) d \mathbf{r'} d ct
\end{equation}
From verial theorem \cite{Lif} we can conclude that the first part of free energy \ref{FERT} are average value of spur tensor energy - momentum of the free particles $\int \langle T_{ii} \rangle  d (\mathbf{r}) $.  In relativistic theory \cite{Lif}, \cite{Zel} the tensor energy -momentum for macroscopic system must take the form
\begin{equation}
T_{i,j}=(\epsilon + P)u_{i}u_{j}-P\delta_{ij}
\end{equation}
where $u_{i}$ four-velocity with condition $u_{i}u^{i}=1$, $\epsilon$ is energy density and $P$ is pressure in system. As shown in \cite{Lif}, \cite{Zel} the first part in action  determine spur of tensor energy - momentum  $T=Sp T_{ij}=3P+\epsilon $ in four dimensional Euclidean space which can rewrite in different curved four dimensional space take into account of volume element  $d \Omega =\sqrt{-g} d^4x $ with  $g=\det g_{ij} $ where metric tensor $g_{ij}$. After that we should be take into account the part of energy which is spent on the distribution of matter, which forms the geometry \cite{Rebesh}: 
\begin{equation}
S_{g}=\frac{1}{c} \int W(\mathbf{r},\mathbf{r'})\rho(\mathbf{r})\rho(\mathbf{r'}) d \mathbf{r'} d ct \simeq -\frac{c^2}{32 \pi G }\int R d \Omega
\end{equation}
where $R$ is curvature of space and $d \Omega =\sqrt{-g} d^4x $ - standard form of element four dimensional volume $g=\det g_{ij} $ with metric tensor $g_{ij}$. Variation of action by metric tensor give the equation for gravitational field if assume that the curvature are induced the distribution of mater with energy-moment tensor $T_{ij}$. 
In according general theory of relativity the Einstein equation for curvature can obtain from Hilbert variation principle the minima of this action together with action of matter we can obtain the Einstein equation in the well - known form
\begin{equation}
R_{ij}= \frac{8 \pi G}{c^4}( T_{ij}-\frac{1}{2}Tg_{ij})
\end{equation}
This presentation of equation of gravitational field are not full, because without tensor second rank $R_{ij}$ are present and another tensor second rank as $g_{ij}$. Simple linear theory can include the simplistic linear combination in the form $R_{ij}+ \alpha g_{ij}= \beta T_{ij}$. From such simple reason we can take into account the possible part which proportional covariant conserve symmetric tensor. Such native tensor in this case are metric tensor. This possibility realize in Einstein equation with cosmological constant $\Lambda = V(\phi)$ \cite{Lin}. From this equation we can conclude that in dynamic of Universe crucial role play only gravitational force, any other forces in such dynamic not take part. We can note again, that the geometry determine distribution of matter. Without this distribution we can not say about any geometry.

\textbf{Conclusion}

In general case the interacting particle systems are non-equilibrium a priory. Indeed, the up-to-date statistical description of non-equilibrium systems considers only probable structures in a interacting systems, but does not describe metastable states  A new approach that employs of a non-equilibrium statistical operator with allowance for inhomogeneous distributions of particles has beenproposed. This method employs the saddle-point procedure to find the dominant contributions to the partition function and allows us to obtain all the thermodynamic parameters of the system. This approach makes it possible to solve the problem with inhomogeneous distributions of particles. On much longer timescales, an evolution towards the true thermal equilibrium is postulated. In this way we can solve the complicated problem of statistical description of systems with garvitational interaction.

We made an attempt to find motivations for the Universe evolution.  The changing in the local thermodynamic potential induce the inhomogeneous distribution of matter is used to determination its distribution and to determine the geometry of this distribution. The statistically equilibrium distribution of matter fully determine the geometry. This interpretation of geometry makes it possible to follow the evolution of the system as a dynamical geometry. The presented article answers the question  from what it possible to substantiate the general relativistic equations  on the basis of statistical methods for describing the behavior of systems.
\subsection{Acknowledgment}: This work was partially supported by the Target Program of Fundamental Research of the Department of Physics and Astronomy of the National Academy of Sciences of Ukraine "Mathematical models of nonequilibrium processes in open systems"
(N0 120U100857).


\begin{thebibliography}{99}

\bibitem{Tir} W. Thirring, Z. Phys. \textbf{235}, 339,  (1970).

\bibitem{Cav} Pierre-Henri Chavanis, Carole Rosier, and Clément Sire, Physical Review E, \textbf{66}, 036105 (2002)

\bibitem{Zub}  D. N. Zubarev, Non-equilibrium statistical thermodynamics (Consultans Bareu, New York), (1974)
 
\bibitem{Lal} Victor Laliena, Nuclear Physics B 668,403 (2003)

\bibitem{Pak} Renato Pakter, Bruno Marcos and Yan Levin, Physical Review Letters,\textbf{111}, 230603 (2013), 

\bibitem{Ben} Fernanda P. C. Benetti, Ana C. Ribeiro-Teixeira, Renato Pakter, and Yan Levin, Physical Review Letters.\textbf{113}, 100602 (2014)

\bibitem{Lev1}  B. I. Lev, International Journal of Modern Physics B, {\bf \ 25}, 2237 (2011).

\bibitem{Lev2} B.I. Lev, Journal of Modern Physics, \textbf{9}, 2223 (2018)

\bibitem{Lev3} B.I. Lev, Journal of Modern Physics, \textbf{10}, 687 (2019) 

\bibitem{Kle}  H. Kleinert , Gauge Field in Condensed Matter, Word Scientific, Singapure,(1989).

\bibitem{Hubbard}  J. Hubbard, Phys. Rev. Lett. \textbf{3}, 77 (1959)

\bibitem{Str}  R.L. Stratonovich, Sov. Phys. Dokl. \textbf{2}, 416 (1958); J. Hubbard, Phys. Rev. Lett. \textbf{3}, 77 ( 1958).

\bibitem{Fey}  R. P. Feynman, Statistical Mechanics, California Institute of Technology,(1972).

\bibitem{Lif}  L. D. Landau and E. M. Lifshiz, Field theory (Pergamon, London), (1981)

\bibitem{Gor}  S. B. Goryachev, Phys. Rev. Lett, \textbf{72}, 1850, (1994)

\bibitem{Lev}  B. I. Lev and A. Ya. Zhugaevych, Phys. Rev. E.{\bf \ 57, } 6460 (1998)

\bibitem{Lev4}  B. I. Lev and A. G. Zagorodny, Phys. Rev.E, {\bf \ 84}, 061115 (2011)

\bibitem{Zel} Zel'dovich, Ya. B.and  Novikov, I. D, ''{\it Theory of gravitation and the evolution of stars}'', Nauka,  Moskva  (1971)

\bibitem{Hilb} D. Hilbert, Die Grundlagen der Physik (Erste Mitteilung), Koeniglichen Ges. Wiss. Goett., Math.-Phys. Kl. Nachr. (1915) 395.

\bibitem{Ein} A. Einstein, Ann. Phys. 354 (1916) 769.

\bibitem{Wein} S. Weinberg, Rev. Mod. Phys. 46 (1974) 255.

\bibitem{Wu} T.T. Wu, C.N. Yang, in: H. Mark, S. Fernbach (Eds.), Properties of Matter under Unusual Conditions, Wiley–Interscience, New York, 1969, 349pp.

\bibitem{Rebesh} A. P. Rebesh , B. I. Lev , Phys.Lett.A,{\bf 381}, 2538, (2017) DOI:o5.0510375-9601

\bibitem{Lin}  A. D. Linde, ''{\it Elementary particle physics and inflationary cosmology}'', Horwood Academic. Switzerland, (1990)

\bibitem{Chan} S. Chandrasekhar, An introduction to the study of stellar structure. New York: Dover Publications,(1942).



\end{thebibliography}
\end{document}